\RequirePackage[2020-02-02]{latexrelease}
\documentclass[sigconf]{acmart}

\usepackage{fancyhdr}
\usepackage{url}
\usepackage{microtype}
\usepackage{algorithmic}
\usepackage{textcomp}
\usepackage{xcolor}
\usepackage{enumitem}
\usepackage{tikz}
\usepackage{siunitx}
\usepackage{soul}
\sethlcolor{yellow}
\usepackage{hyperref}

\usepackage{titlesec}
\titleformat*{\paragraph}{\bfseries}
\titlespacing*{\paragraph}{0pt}{1ex plus .1ex minus .2ex}{2pt}

\usepackage{setspace}
\usepackage{def}

\AtBeginDocument{%
  \providecommand\BibTeX{{%
    \normalfont B\kern-0.5em{\scshape i\kern-0.25em b}\kern-0.8em\TeX}}}

\usepackage{graphicx}
\usepackage{float}
\usepackage{rotating}
\usepackage{comment}
\usepackage{subfig}
\usepackage{makecell}
\usepackage{multirow}
\usepackage{subfloat}
\usepackage{balance}

\definecolor{mygreen}{rgb}{0,0.6,0}
\definecolor{mygray}{rgb}{0.5,0.5,0.5}
\definecolor{mymauve}{rgb}{0.58,0,0.82}

\usepackage[ruled]{algorithm2e} 

\SetAlFnt{\small}
\SetAlCapFnt{\small}
\SetAlCapNameFnt{\small}
\SetAlCapHSkip{0pt}
\IncMargin{-\parindent}

\begin{document}

\title[]{UniHeap: Managing Persistent Objects \\Across Managed Runtimes for Non-Volatile Memory\footnotemark[1]}



\author{Daixuan Li, Benjamin Reidys, Jinghan Sun, Thomas Shull, Josep Torrellas, Jian Huang}

\authornote{This work has been published at SYSTOR'21 \cite{uniheap:systor2021}.}

\affiliation{%
  \institution{University of Illinois Urbana-Champaign}
  \streetaddress{Address}
  \city{}
  \state{}
  \country{}}

\renewcommand{\shortauthors}{} 

\begin{abstract}
    \vspace{-0.5ex}
\looseness=-1
Byte-addressable, non-volatile memory (NVM) is emerging as a promising technology. 
To facilitate its wide adoption, 
employing NVM in managed runtimes like JVM has proven to be an effective approach (i.e., managed NVM).
However, such an approach is runtime specific, which lacks 
a generic abstraction across different managed languages. 
Similar to the well-known filesystem primitives that allow diverse programs 
to access same files via the block I/O interface, 
managed NVM deserves the same system-wide property for 
persistent objects across managed runtimes with low overhead. 

In this paper, we present \pname{}, a new NVM framework for managing persistent 
objects. 
It proposes a unified persistent object model 
that supports various managed languages, and manages NVM within a shared heap 
that enables cross-language persistent object sharing. \pname{} 
reduces the object persistence overhead by managing the shared heap in a log-structured 
manner and coalescing object updates during the garbage collection. 
We implement \pname{} as a generic framework and extend it to 
different managed runtimes that include HotSpot JVM, cPython, and JavaScript engine 
SpiderMonkey. 
We evaluate \pname{} with a 
variety of applications, such as key-value store and transactional 
database. 
Our evaluation shows that \pname{} 
significantly outperforms state-of-the-art object sharing approaches, 
while introducing negligible overhead to the managed runtimes. 

\end{abstract}

\maketitle

\vspace{-2ex}	
\section{Background and Motivation}
\vspace{-0.5ex}
\label{sec:intro}
Non-volatile memory (NVM), such as phase-change memory (PCM), 
resistive RAM (ReRAM), NVDIMM, and Intel DC persistent memory, 
has become a promising technology that offers near-DRAM speed, scalable storage capacity, and data durability. 
To facilitate its wide adoption in practice, 
its management and use in software systems have attracted much attention recently. 

Specifically, 
many NVM frameworks and libraries 
have been developed, such 
as Mnemosyne, NVHeaps, and Intel PMDK. 
However, most of them require developers to explicitly specify the persistent data 
structures in their programs, which significantly increases the development burden. 
To address this issue, recent researches proposed to integrate NVM into managed runtimes 
like JVM, 
in which they leverage the runtime system to transparently manage objects 
in NVM. 
As the managed languages, such as Java, Python, and JavaScript, have become the most 
popular programming languages, 
such an approach is becoming pervasive. We define this approach 
as \textit{managed NVM} in this paper.  

Although utilizing managed runtimes to use NVM has proven to be an effective approach  
to simplify the NVM programming,  
state-of-the-art approaches are runtime specific, and lacking an important system-wide property 
-- \textit{persistent object management across managed runtimes}. 
It is not easy for a 
Python program to directly access an object persisted by a Java program, 
as their runtime-specific object format and layout are different. 

As system-wide shared resource, managed NVM deserves data sharing,  
and it provides non-volatility as shared persistent storage does.
Similar to the file systems developed for persistent storage, which manages data in the format of files, and allow different programs 
to access shared files with block I/O interface, 
it is highly desirable to enable diverse managed languages to access shared persistent 
objects efficiently, which could pave the way for developing managed NVM into a generic 
approach.
 
As more applications are developed based on managed runtimes today,  
they usually use different runtime instance for each individual component. 
And platform operators also prefer to deploy multiple 
runtime instances on the same machine to best utilize the compute and memory 
resources. 
Take the web service for example, its frontend uses JavaSript runtime while its backend adopts 
JVM. 

To achieve persistent object sharing, a straightforward approach is to leverage the persistence 
layer available in managed runtimes to persist objects to file systems or database.  
Typical examples include Java Persistence API (JPA) and Java 
Data Objects (JDO). 
However, 
they cause significant performance overhead.
Wegiel et al. proposed to exploit the shared memory to enable the 
cross-language cross-runtime communication, unfortunately, it does not support NVM.  
The industry has developed Thrift and Protocol Buffers
to facilitate the interoperation 
across multiple languages, however, they suffer from significant marshalling and unmarshalling 
overheads.



\begin{figure}[t]
	\centering
	\includegraphics[scale=0.72]{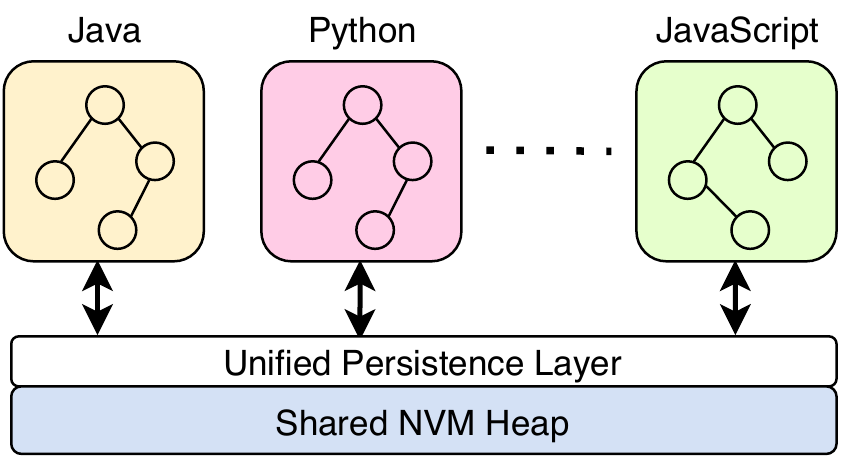}
	\vspace{-2ex}
	\caption{Overview of \pname{}}
	\vspace{-3.5ex}
	\label{fig:overview}
\end{figure}



\vspace{-2ex}	

\begin{table}[t]
    \caption{The mapping of language types in \pname{}.}
    \label{tab:type}
    \vspace{-2.5ex}	
    \centering
    \scriptsize
    \begin{tabular}{|p{32pt}<{\raggedleft}|p{22pt}<{\centering}|p{15pt}<{\centering}|p{12pt}<{\centering}|p{12pt}<{\centering}|p{12pt}<{\centering}|p{18pt}<{\centering}|p{28pt}<{\centering}|}
	\Xhline{0.9pt}		
       	\textbf{\makebox[25pt][c]{Java}} & boolean, byte & char & int & long & float & double & reference, array \\
       	\hline
       	\textbf{\makebox[25pt][c]{Python}} & - & - & int & long & float & - & list, dict, tuple \\
        \hline
       	\textbf{\makebox[27pt][c]{JavaScript}} & boolean & - & num & num & num & num & array \\
	\Xhline{0.9pt}
		\textbf{UniHeap} & char & short & int & long & float & double & reference \\
       	\Xhline{0.9pt}
    \end{tabular}
    \vspace{-5ex}	
\end{table}

\section{Design and Implementation}
\label{sec:design}
\vspace{-0.5ex}

In this paper, we develop a lightweight NVM framework, named \pname{}, for persistent object 
management across a diversity of managed runtimes(see Figure~\ref{fig:overview}).
\pname{} has a unified persistence layer located between the upper-level runtimes 
and the underlying managed NVM heap. 

\textbf{Unified Persistence Layer.}
We design a unified persistence layer (UPL) in \pname{} for two purposes. 
First, UPL has a language-neutral object model, such that it can be extended to 
support new managed languages. Second, UPL should facilitate object persistence 
for managed runtimes to achieve low persistency overhead.   
Unlike recently proposed PCJ for NVM, \pname{} does not introduce new 
type system. 
\pname{} provides two built-in types: \emph{numeral} type  
and \emph{reference} type. It does not provide container types, such as \textit{list}, \textit{dict}, and 
\textit{tuple} in Python, 
as developer can implement their own container type based on these built-in types. 
As shown in Table~\ref{tab:type}, the numeral type includes \emph{char}, \emph{short}, 
\emph{int}, \emph{long}, \emph{float}, and \emph{double}. 
For the reference type, the object field stores the pointer to other persistent objects. 
It is worth noting that \textit{array} is also treated as an object in \pname{}. 
Thus, its object field can store a pointer to an array. 
As we provide a transparent type system for managed runtimes, different managed language 
needs to map their type into the type system of \pname{}. We show the mapping of the popular managed 
languages that include Java, Python, and JavaScript in Table~\ref{tab:type}. 
As for Object persistence,
we manage to stay compatible with the data persistence approaches in the existing NVM 
frameworks.

\begin{figure}[t]
	
	\centering
	\includegraphics[scale=0.62]{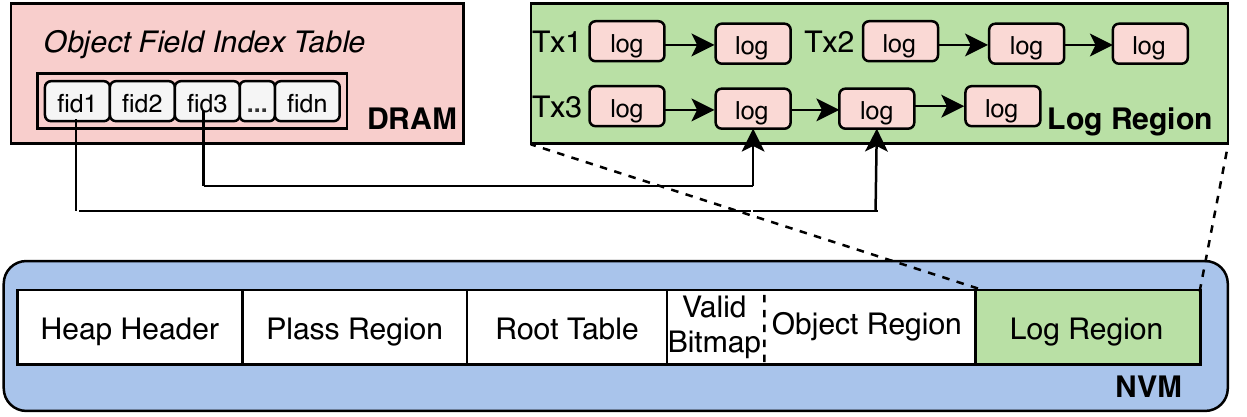}
	\vspace{-2ex}
	\caption{The shared NVM heap structure in \pname{}.}
	\label{fig:nvmheap}
	\vspace{-5ex}
\end{figure}


\textbf{Shared NVM Heaps.} \pname{} stores persistent objects in shared NVM heaps. 
It organizes the shared NVM heap into five regions as shown in Figure~\ref{fig:nvmheap}: 
heap header, plass region, root table region, 
object region, and log region. The heap header stores the metadata for the corresponding heap, 
including the heap name (32 bytes) and heap size (8 bytes). The plass region stores all the class descriptors 
for \pname{} objects. The root table stores all the durable root objects. Each root is a key-value pair 
with the format of $<$$root\_name, root\_addr$$>$. The object region contains an object valid bitmap, 
which is used by the GC for reclaiming persistent objects in \pname{}. 
Its remaining part is organized into numerous fixed-size (16 bytes) chunks, 
each of which stores an object header. The log region stores the object updates in a 
{log-structured manner}.  

\looseness=-1
To facilitate out-of-place update, 
we decouple the object header from its data, 
based on the insight that object header is not frequently updated in a transaction. 
We store the object header in the object region, and object data in the log region. 
Once a managed runtime allocates an object from the shared NVM heap, \pname{} will bump the pointer in the object region 
to allocate an object header.

The out-of-place update approach can reduce write traffic to NVM, 
however, 
it requires address remapping to retrieve the latest data for read operation. 
To address this challenge, \pname{} employs a \emph{per-object} mapping table for address 
translation. 
Since each object field has a fixed index value during the object lifetime, \pname{} use the 
field index to store the address of the latest updates in the log region. 
The address translation procedure is efficient ($O(1)$). In addition, \pname{} caches 
the object field index table in the fast DRAM for further performance improvement (see Figure~\ref{fig:nvmheap}). 

\looseness=-1
\textbf{Persistent Object Sharing.}
\pname{} provides a set of interfaces for managed runtimes  
to access the shared NVM heap, 
we classify them into two categories: language-neutral and language-related. 
The language-neutral interfaces do not need support from managed runtimes. 
For example, \pname{} implements the \emph{alloc\_obj} interface with its own 
object allocation policy, 
which is independent from the managed runtime. 
For those language-related interfaces that include only \emph{init\_plass} and \emph{exists\_plass}, 
they require runtimes to have their own implementations. For instance, when \pname{} 
initializes a class with \emph{init\_plass},  
\pname{} requires runtime support to fill up the plass structure,  
since the class metadata 
is stored in the language source file.

\textbf{Coordinated GC for Persistent Objects.}
\pname{} uses the mark-and-compact GC to reclaim persistent objects. As managed runtimes use \pname{} to store their persistent objects, 
\pname{} will track   
their references and reachability from runtimes.  \pname{} is 
able to track each runtime's own reference to a persistent object efficiently, can coordinate with 
multiple runtimes to reach a system-wide safety point, and then stop the world to perform 
the GC.  The GC of \pname{} is crash safe by ensuring each GC phase is idempotent.  Similar to previous work, the GC of \pname{} consists of four phases: (1) marking phase, 
(2) relocation phase, (3) compaction phase, and (4) clean-up phase.
The marking phase is naturally idempotent, since this phase does not change the heap states. 
As for the compaction phase, we maintain the old object region until the clean-up phase.  
Therefore, upon a crash or failure, \pname{} can redo the GC during the system recovery.


Overall, we make the following contributions in this paper.  

\vspace{-1ex}	
\begin{itemize}[leftmargin=*]

\item We propose a generic NVM framework that provides a unified 
	object model to enable efficient persistent object sharing cross diverse managed 
	runtimes within NVM. 

\vspace{0.5ex}
\item We present a shared NVM heap for managing persistent objects.  
	It manages objects in a log-structured manner and supports  
	both in-place and out-of-place updates to reduce data persistence overhead, 
	while ensuring the crash-safety. 


\vspace{0.5ex}
\item We develop an efficient GC scheme by decoupling the metadata and data of persistent 
	objects in the NVM heap, and coordinate GC operations with managed runtimes 
		to ensure the correctness of object cleanups. 

\vspace{0.5ex}
\item We enable \pname{} to  
	support three popular
	managed runtimes, including HotSpot JVM, cPython, and JavaScript. 

\vspace{-0.5ex}
\end{itemize}

\vspace{-2ex}	
\section{Evaluation}
\vspace{-1ex}

\begin{figure}[t]
	\centering
	\includegraphics[width=0.95\linewidth]{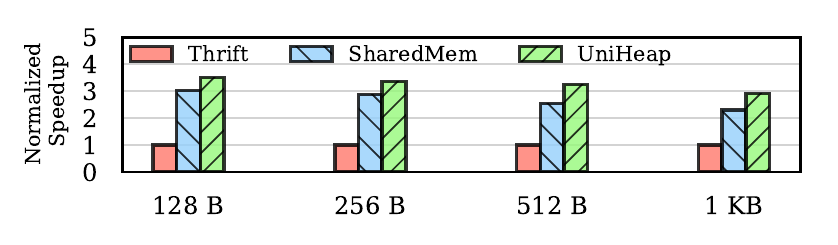}
	\vspace{-4ex}
	\caption{Performance comparison of different persistent object sharing approaches. }
	\label{fig:sharing}
	\vspace{-2.5ex}
\end{figure}

\begin{figure}[t]
	\vspace{-1ex}
	\centering
	\includegraphics[width=0.95\linewidth]{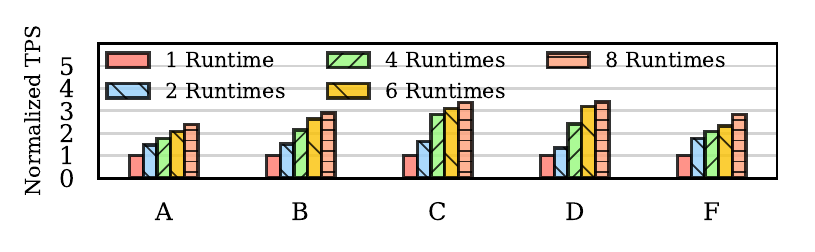}
	\vspace{-4ex}
	\caption{Scalability of \pname{}.}
	\label{fig:scalability}
	\vspace{-2.5ex}
\end{figure}

\begin{figure}[t]
	\vspace{-1ex}
	\centering
	\includegraphics[width=0.95\linewidth]{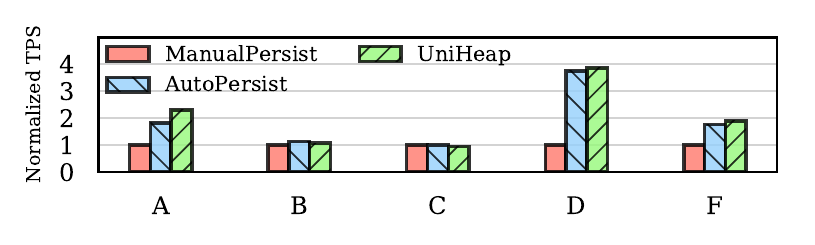}
	\vspace{-4ex}
	\caption{Performance of running HotSpot JVM with \pname{}, when using QuickCached+YCSB.}
	\label{fig:ycsb}
	\vspace{-5ex}
\end{figure}

We implement UniHeap system prototype as shared library with 9,163 lines of C programming code. As for the shared NVM heap, UniHeap uses the memory-mapped interface with Direct Access (DAX) enabled for fast access to the NVM device.
To evaluate the efficiency of \pname{}, we run a variety of data-intensive applications and typical 
benchmarks for different managed runtimes, including Yahoo Cloud Service Benchmarks (YCSB)
for Java, Python Performance Benchmark Suite for Python, and JetStream2
for JavaScript. 
\label{sec:eval}
Our evaluation(see Figure~\ref{fig:sharing},Figure~\ref{fig:scalability},and Figure~\ref{fig:ycsb})  shows that 
(1) \pname{} performs better than state-of-art approaches of persistent object sharing; (2) It can scale the persistent object sharing 
as we increase the number of managed runtimes;  
(3) \pname{} enables persistent object sharing 
across different runtimes without introducing much performance 
overhead. We presented the detailed evaluation in \cite{uniheap:systor2021}.

\vspace{-2ex}	
\bibliographystyle{plain}
\bibliography{ref,references,pldiref}

\begin{thebibliography}{1}

\bibitem{uniheap:systor2021}
Daixuan Li, Benjamin Reidys, Jinghan Sun, Thomas Shull, Josep Torrellas, and
  Jian Huang.
\newblock Uniheap: Managing persistent objects across managed runtimes for
  non-volatile memory.
\newblock In {\em Proceedings of the 14th ACM International Conference on
  Systems and Storage}, SYSTOR '21, New York, NY, USA, 2021. Association for
  Computing Machinery.

\end{thebibliography}

\end{document}